\documentclass[reprint,preprintnumbers,amsmath,amssymb,floatfix,superscriptaddress,prl,nobibnotes,showpacs]{revtex4-1}
\usepackage{graphicx}
\usepackage{bm}
\usepackage{hyperref}
\usepackage{placeins}

\begin{document}

\title{Resonant X-ray Scattering Measurements of a Spatial Modulation of the Cu~$3d$ and O~$2p$ Energies in Stripe-Ordered Cuprate Superconductors} 

\author{A. J. Achkar}
\affiliation{Department of Physics and Astronomy, University of Waterloo, Waterloo, N2L 3G1, Canada}
\author{F. He}
\affiliation{Canadian Light Source, University of Saskatchewan, Saskatoon, Saskatchewan, S7N 0X4, Canada}
\author{R. Sutarto}
\affiliation{Department of Physics and Astronomy, University of British Columbia, Vancouver,V6T 1Z4, Canada}
\author{J. Geck}
\affiliation{Leibniz Institute for Solid State and Materials Research IFW Dresden, Helmholtzstra\it{\ss}e 20, 01069 Dresden, Germany}
\author{H. Zhang}
\affiliation{Department of Physics, University of Toronto, Toronto, M5S 1A7, Canada}
\author{Y.-J. Kim}
\affiliation{Department of Physics, University of Toronto, Toronto, M5S 1A7, Canada}
\author{D. G. Hawthorn}
\affiliation{Department of Physics and Astronomy, University of Waterloo, Waterloo, N2L 3G1, Canada} 


\begin{abstract}
A prevailing description of the stripe phase in underdoped cuprate superconductors is that the charge carriers (holes) phase segregate on a microscopic scale into hole rich and hole poor regions. We report resonant elastic x-ray scattering measurements of stripe-ordered La$_{1.475}$Nd$_{0.4}$Sr$_{0.125}$CuO$_4$ at the Cu $L$ and O $K$ absorption edges that identify an additional feature of stripe order. Analysis of the energy dependence of the scattering intensity reveals that the dominant signature of the stripe order is a spatial modulation in the energies of Cu $3d$ and O $2p$ states rather than the large modulation of the charge density (valence) envisioned in the common stripe paradigm. These energy shifts are interpreted as a spatial modulation of the electronic structure and may point to a valence-bond-solid interpretation of the stripe phase.
\end{abstract}
\maketitle

Static stripe order in cuprates was first theoretically predicted by mean-field Hubbard model calculations \cite{Zaanan89,Poilblanc89,Schulz89,Machida89} and subsequently observed in lanthanum-based cuprates by neutron and x-ray diffraction\cite{Tranquada95,Tranquada96, Zimmermann98, Kim08, Hucker11, Abbamonte05, Fink09}.   Although still a matter of debate, more recent work has indicated that stripe-like density wave order is generic to the cuprates \cite{Hanaguri04,Kohsaka07,LeBoeuf07,Daou10, Vojta09,Wu11} and plays a significant role in competing with or possibly causing superconductivity \cite{Berg09}.

Microscopically, stripes in the cuprates have been widely described as rivers of charge---hole-rich antiphase domain walls that separate undoped antiferromagnetic regions.  However, alternate models with different underlying physics, such as the valence bond solid (VBS), have also been proposed to explain stripe order \cite{Read89,Sachdev03,Vojta99}.  VBS models involve singlet formation between neighbouring spins and, in contrast to other models of stripe order, may occur with a small modulation of the charge density \cite{Sachdev03}.

Distinguishing which of these models is most relevant to stripe order in the cuprates is challenging since the models share many symmetries and experimental signatures.  In particular, direct evidence for charge-density modulations, which may distinguish various models, has been elusive.  Neutron and conventional x-ray scattering are only sensitive to lattice displacements.   It is therefore only inferred indirectly that these lattice displacements are induced by modulations in charge density (valence).  Resonant soft x-ray scattering (RSXS) offers a means to couple more directly to modulations in the electronic structure, including charge density modulations.  By performing an x-ray diffraction measurement on resonance (at an x-ray absorption edge), the atomic scattering form factor $f(\omega)$ is enhanced and made sensitive to the valence, orbital orientation and spin state of specific elements.  A key feature of RSXS is that the energy dependence of the scattering intensity through an absorption edge differs for lattice distortions, charge-density modulations or other forms of electronic ordering, providing a means to distinguish these different types of order.  

In the cuprates, RSXS of the [2$\varepsilon$, 0, $L$] charge density wave (CDW) superlattice peak has been measured in stripe-ordered La$_{2-x}$Ba$_x$CuO$_4$ (LBCO)  \cite{Abbamonte05}, La$_{2-x-y}$Eu$_y$Sr$_x$CuO$_4$~(LESCO)\cite{Fink09,Fink11} and  La$_{1.475}$Nd$_{0.4}$Sr$_{0.125}$CuO$_4$ (LNSCO)~\cite{Wilkins11} at the O $K$ (1$s \rightarrow 2p$) and Cu $L$ (2$p \rightarrow 3d$) absorption edges, which provide sensitivity to the O 2$p$ and Cu 3$d$ orbitals that are central to the physics of the cuprates. These measurements have been interpreted as direct evidence for a large valence modulation on the O sites \cite{Abbamonte05}. Moreover, it is argued that a modulation of the valence occurs primarily on the O sites and not on the Cu sites, which are instead subject to lattice distortions induced by the valence modulation on the O sites \cite{Abbamonte05,Fink09}. However, efforts to model the energy dependence of the scattering intensity based on this picture are not truly reconciled with experiment, leaving this interpretation open to question \cite{Fink09}.

In this Letter, we present O $K$ and Cu $L$ edge RSXS measurements of LNSCO.  The energy dependence of the scattering intensity is modelled using x-ray absorption measurements to determine $f(\omega)$ at different sites in the lattice, a procedure that has proven effective in describing the scattering intensity of valence modulations in the chain layer of ortho-II YBa$_2$Cu$_3$O$_{6+\delta}$ (YBCO) \cite{Hawthorn11b}. Contrary to previous analysis of LESCO \cite{Fink09} and LBCO \cite{Abbamonte05}, we show that the resonant scattering intensity is best described by small energy shifts in the O $2p$ and Cu $3d$ states at different Cu and O sites rather than a valence modulation of O and a lattice displacement of Cu.

\begin{figure*}[htb]
\centering
\resizebox{6.6in}{!}{\includegraphics{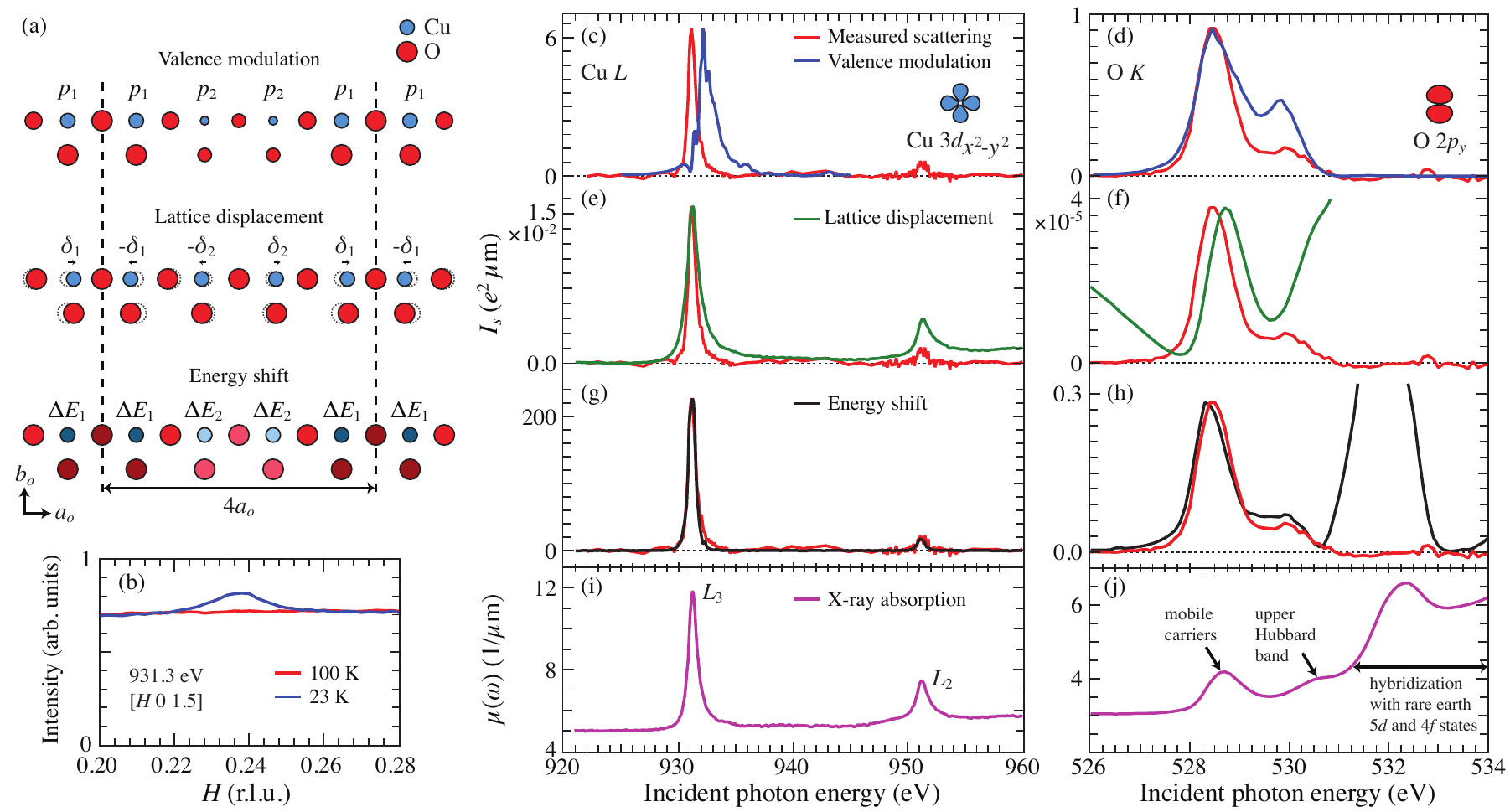}}
\caption{(color online). {Resonant scattering of 1/8 doped LNSCO at the Cu $L$ and O $K$ edges.}  (a) Schematic representations of bond-centred stripe ordering for the three models proposed to describe the resonant scattering energy dependence.  (b) $H$ scan through the CDW superlattice peak at [$H$, 0, 1.5] and at the peak of the Cu $L_3$ absorption edge \cite{Hawthorn11b}. (c)---(h) Scattering intensity as a function of photon energy through the Cu $L_{3,2}$ and O $K$ absorption edges. The measured intensity (red line) is compared to the scattering intensity of valence modulation (blue line), lattice displacement (green line) and energy shift (black line) models of the stripe-ordering. The best agreement with experiment is for the energy shift model. (i) and (j) X-ray absorption $\mu(\omega)$ at the Cu $L_{3,2}$ (i) and O $K$ (j) absorption edges measured using total electron yield.}
\label{fig1}
\end{figure*}

The measured intensity for scans through ${\bf Q}$ = [$H$, 0, 1.5] with the photon energy at the peak of the Cu $L_3$ absorption edge (931.3 eV) are presented in Fig.~1(b) (see Supplemental Material \cite{supp} for methods).  Below the stripe-ordering transition temperature of $\sim$70 K, a clear superlattice peak is observed at ${\bf Q} = [2\varepsilon, 0, L]=[0.236, 0, 1.5]$.  Above the stripe-ordering transition temperature, a large, smoothly varying fluorescence background is observed.  To determine the photon energy dependence of the scattering intensity, $H$ scans through the superlattice peak were performed at 22 K as a function of photon energy at $L$ = 1.5 for Cu and $L$ = 0.2 for O.  The fluorescence background is fit and subtracted from each scan.  The integrated intensity of the superlattice peak is then determined as a function of photon energy for the O $K$ and Cu $L$ edges [Figs.~1(c) and 1(d)]. The resulting spectra are qualitatively similar to previous measurements on LBCO \cite{Abbamonte05} and LESCO \cite{Fink09}.    Importantly, our measurements extend the previous Cu $L$ edge measurements to include the $L_2$ edge, which proves valuable in distinguishing models for the stripe phase.  An important feature of our measurement is that all scattering measurements are performed with the incident x-ray polarization along the $b_o$ axis of the sample.  As a result, the scattering intensity will be sensitive to only the O $p_y$ and not the O $p_x$ orbitals. Assuming doped holes go only into $\sigma$-bonded orbitals of O, this measurement geometry is only sensitive to half of the oxygen atoms; the site-centred and not the bond-centred oxygen.  This fact simplifies the expression for the structure factor.

The measured energy dependence of the scattering intensity is compared to three model calculations [see Fig.~\ref{fig1}(a)]:  1. valence modulation, a spatial modulation in the valence of the Cu and O;  2. lattice displacement, a small displacement  of the Cu and O atoms from their equilibrium positions outside the stripe-ordered phase; and 3. energy shift, a spatial modulation in the energy of the Cu $3d$ and O $2p$ states.  The first two models essentially follow previous analysis of RSXS in LBCO and LESCO \cite{Abbamonte05, Fink09}.

The three models differ in the structure factor (described in the Supplemental Material \cite{supp}) and the energy dependence of the atomic scattering form factor $f(\omega)$.  These two factors give rise to a different energy dependence to the scattering intensity, $I_s(\omega)$.  For the valence modulation model, $I_s(\omega) \propto |f(\omega,p_2)-f(\omega,p_1)|^2/\mu(\omega)$, where $p_1$ and $p_2$ are the local hole concentrations (valence) at different sites [see Fig.~\ref{fig1}(a)] and $\mu(\omega)$ is the absorption coefficient.  For the lattice displacement model, $f(\omega)$ is the same at each site for a given element and $I_s(\omega) \propto |f(\omega)|^2/\mu(\omega)$.  Finally, for the energy shift model,  $I_s(\omega) \propto |f(\hbar \omega + \Delta E)-f(\hbar \omega -\Delta E)|^2/\mu(\omega)$, similar to the valence modulation model but with an energy shift $\pm \Delta E$ at different sites instead of a modulation in valence.  In all three models, the site specific $f(\omega,p_{1,2},\Delta E)$ are determined from x-ray absorption measurements.  

\begin{figure}[b]
\centering
\resizebox{89mm}{!}{\includegraphics{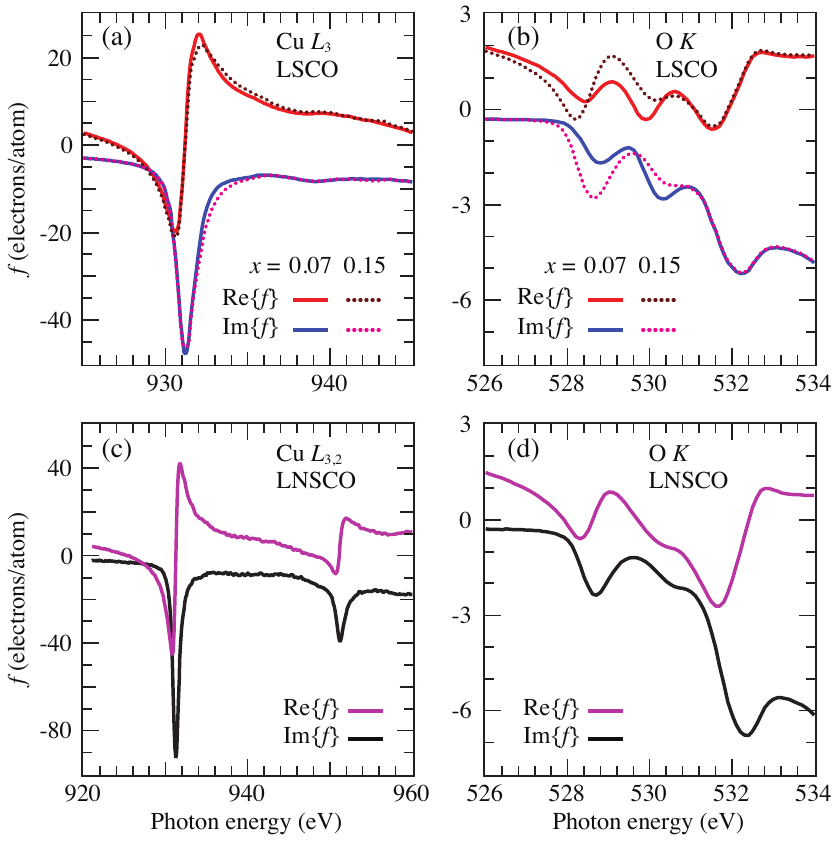}}
\caption{(color online). The atomic scattering form factors $f(\omega)$ as a function of photon energy through the Cu $L$ and O $K$ absorption edges for (a), (b) La$_{2-x}$Sr$_{x}$CuO$_4$ ($x$ = 0.07 and 0.15)  and (c), (d) La$_{1.475}$Nd$_{0.4}$Sr$_{0.125}$CuO$_4$. $f(\omega)$ for LSCO is determined using XAS measurements from Ref.~\cite{Chen92}. 
}
\label{fig2}
\end{figure}

{\it Valence modulation model.}---For the valence modulation model, x-ray absorption spectroscopy (XAS) on samples with different doping levels from Ref.~\cite{Chen92} are used to determine $f(\omega, p_{1,2})$.  This procedure found very good agreement between experiment and calculations for a modulation of the Cu valence in the chain layer of oxygen-ordered YBCO \cite{Hawthorn11b}. In lanthanum based cuprates, the key features of the O $K$ edge XAS are two preedge peaks at 528.6 eV and 530.5 eV that are due to hybridization between Cu $3d$ and O $2p$ states and have been assigned to the mobile doped holes and the upper Hubbard band respectively [Fig.~\ref{fig1}(j)] \cite{Chen91,Chen92,Romberg90}.  The intensities of these two peaks evolve strongly with doping, whereas the spectra at higher energy are doping independent and dominated by O $2p$ states hybridized with rare earth $5d$ and $4f$ states \cite{Nucker88}.  As argued in Ref.~\cite{Fink09}, the scattering intensity for a valence modulation of arbitrary magnitude can be modeled using XAS measured at two different dopings. Here $f_j(\omega)$ [Fig.~2(b)] and the scattering intensity expected for a valence modulation [Fig.~1(d)] is calculated from XAS in La$_{2-x}$Sr$_x$CuO$_4$ (LSCO) at $x$ = 0.07 and 0.15 from Ref.~\cite{Chen92}, corresponding to a hole modulation of $\delta p = p_1 - p_2$ = 0.08.  Although this calculation successfully produces two peaks at approximately the correct energies, it strongly overestimates the intensity of the peak at 529.9 eV \cite{Fink09}.  A different choice of doping values to determine $f(\omega,p_{1,2})$ impacts the magnitude of the scattering, scaling it roughly as $\delta p^2$, and produces differences in the line shape (see Supplemental Material \cite{supp}). However, calculations using existing XAS data are all similarly inconsistent with the measured RSXS line shape.  

A similar analysis, again using XAS from Ref.~\cite{Chen92} to determine $f(\omega,p)$ [Fig.~2(a)], can be applied to the Cu $L$ edge.  The XAS for the Cu $L$ edge exhibits two primary peaks at 931.3 eV and 951.3 eV corresponding to the $L_3$ and $L_2$ edges that are split by the spin-orbit coupling of the 2$p$ core electrons.   Focusing on the $L_3$ edge, the XAS is comprised of a peak (931.3 eV) and a shoulder (932.3 eV) that are associated with $d^9$ (a single hole in the $d_{x^2-y^2}$ orbital) and $d^9\underline{L}$ (doped holes that are primarily on the oxygen ligands) ground states.  Consistent with this assignment, the high energy shoulder evolves much more strongly with doping than the $d^9$ peak \cite{Chen92}.  It follows that the predicted scattering intensity for a valence modulation of the Cu is peaked at the shoulder and not the peak of the XAS [Fig.~1(c)].  As discussed in Refs.~{\cite{Abbamonte05,Fink09}, this is in poor agreement with the energy dependence of the resonant scattering, which is peaked at the maximum of the XAS.  

{\it Lattice displacement model.}---Calculations based on the lattice displacement model [Figs.~1(e) and 1(f)] are also in poor agreement with experiment [here using XAS on our sample of LNSCO in Figs.~1(i) and 1(j) to determine $f(\omega)$ in Figs.~2(c) and 2(d)].   The lattice displacement model at the O $K$ edge predicts large scattering intensity above and below the absorption edge that is not observed in experiment and, at the Cu $L$ edge, scattering intensity that is broader in energy and has a smaller ratio of the $L_3$ to $L_2$ peaks than the measurement.  The calculated magnitude of the scattering intensity assumes a 0.004 \AA ~lattice displacement, as deduced from neutron scattering \cite{Tranquada96}. 

{\it Energy shift model.}---Like the lattice displacement model, XAS on our sample of LNSCO [Figs.~1(i) and 1(j)] is used to determine $f(\omega)$ [Figs.~2(c) and 2(d)] for the energy shift model.  For the two sites (1 and 2), $f(\omega)$ is shifted in energy by $\Delta E = \pm$0.1 eV for both the O $K$ and  Cu $L$ edges.  In contrast to the lattice displacement and valence modulation models, the energy shift model is in very good agreement with experiment.  At the Cu $L$ edge, it captures the correct intensity ratio of the Cu $L_3$ and $L_2$ peaks, the correct width in energy of the scattering and the correct energy position of the maximum of the scattering intensity.  Similarly, at the O $K$ edge, the energy shift model reproduces the energy dependence of the preedge peak. It does not agree with the spectra at higher energy, predicting a large peak at 531.5 eV that is not observed.  However, this apparent discrepancy can be reconciled if we interpret this as evidence that only the low energy states involving hybridization between the O $2p$ and Cu $3d$ states (and not the rare earth $5d$ and $4f$ states) are subject to these energy shifts.   We also note that the choice of $\Delta E$ impacts the magnitude, which scales as  $\Delta E^2$, but not the energy dependence of the calculated scattering intensity, provided $\Delta E$ is less than the energy width of the XAS ($\sim $0.2 eV). As such, $\Delta E$ is neither determined in our analysis, nor should it be viewed as a fitting parameter. 

Our interpretation of the energy shifts is that they are induced by subtle spatial modulations of the local electronic structure. The energy levels of the unoccupied Cu 3$d$ and O 2$p$ states can be described by parameters such as the Cu onsite Coulomb repulsion ($U_{dd}$), the crystal field splitting parameters ($10Dq, Ds, Dt$), the charge transfer energy ($\Delta_{pd}$) and the Cu $3d$---O $2p$ hopping ($t_{pd}$) \cite{Chen91,Hybertsen92,Eskes91}.  Small changes to these parameters can lead predominately to shifts in the Cu $3d$ and O $2p$ energy levels that would manifest themselves as energy shifts in $f(\omega)$ \cite{deGroot05}, unlike the large changes in the spectral weight distribution that are observed with doping \cite{Chen92}. Since the XAS and RSXS are sensitive to the Cu $2p\!\rightarrow\! 3d$ and O $1s\!\rightarrow\! 2p$ transitions, modulations of the Cu $2p$ and O $1s$ binding energies may also contribute to the energy shifts. 

The agreement between the measured scattering intensity and this simple phenomenological model indicates that energy shifts are responsible for the dominant contribution to the resonant scattering intensity.  In comparison, contributions arising from lattice displacements and valence modulations appear to be much less significant.  This is reasonable for the lattice displacement model, given that the magnitude of the calculated scattering intensity is $\sim$4 orders of magnitude weaker than the energy shift model at both the O $K$ and Cu $L$ edges [Since $\Delta E$ is unknown, for this analysis we compare to an arbitrarily chosen value of $\Delta E$ = 100 meV, which serves as a reasonable upper limit value. Comparing to other values of $\Delta E$ involves scaling by $(\Delta E/(100$~meV$))^2$.] 

In contrast, the valence modulation model (assuming $\delta p = 0.08$) is $\sim$4 times larger than the energy shift model at the O $K$ edge.   At the Cu $L$ edge, the valence modulation model is $\sim$35 times weaker than the energy shift model at their respective peak values but is comparable in intensity at 932.3 eV, the peak energy of the valence model calculation.  As such, unlike lattice displacements, we do not expect the energy shift contribution to dominate the scattering intensity for valence modulations of order $\delta p = 0.08$ or larger.  This argues against a large valence modulation, such as those reported in Ref.~\cite{Abbamonte05}, but does not rule out smaller valence modulations. However, even if negligibly small, one can infer that valence modulations must be nonzero, as they must occur for a spatial modulation of the Cu 3$d$ and O 2$p$ energies (i.e., the energy shifts provide indirect evidence for valence modulations).  Placing a precise upper limit on the magnitude of the valence modulation is beyond the scope of the present work, requiring more sophisticated modeling.  However, we note that the energy shift model has the same unoccupied spectral weight, and hence the same valence, for all sites in the stripe phase. As such, the resonant scattering line shapes, which are well described by the energy shift model alone, are consistent with a stripe phase that has a minimal but nonzero valence modulation.

The origin of these modulating energy levels and how they relate to the microscopic mechanism of stripe ordering is an open question. The energy shifts may simply be induced by small charge-density modulations or lattice displacements, yet still be the dominant signature in resonant scattering. Alternatively, they may be a more direct signature of the interactions underlying stripe order. For instance, these energy modulations may point to a VBS description of the stripe phase \cite{Read89,Sachdev03,Vojta99}.  In the VBS picture, stripe order is driven by exchange interactions, which also induce lattice displacements and bond-centered charge order.  However, the magnitude of bond-centered charge density modulations can be small, being screened by long-range Coulomb repulsions.  This may provide an explanation for the lack of clear evidence for a valence modulation from resonant scattering.  In contrast, the energy shifts that we have identified in RSXS may arise naturally out of a VBS, which involves a modulation of the bonding in the lattice. 

Finally, our identification of energy shifts is likely applicable not only to CDW order in other cuprates (energy shifts were recently shown to also describe density wave order in YBCO \cite{Achkar12}), but also to other transition oxides.  For instance, recent first principles calculations have shown that several ``charge'' ordered transition metal oxides exhibit a site dependence to the energies of the electronic states but no site dependence to the total $d$ orbital occupation \cite{Quan12}, similar to our phenomenological energy shift model.

\begin{acknowledgements}
We thank A. Burkov, J. C. Davis, S. Sachdev, E. Fradkin, P. Abbamonte and G. A. Sawatzky for discussions.  This work was supported by CFI, BCSI and NSERC. The Canadian Light Source is supported by NSERC, NRC, CIHR, and the University of Saskatchewan.
\end{acknowledgements}

%

\end{document}